\documentclass[aps,prd,superscriptaddress,tightenlines,twocolumn,nofootinbib,showpacs,floatfix]{revtex4}

\usepackage{graphicx}
\usepackage{dcolumn}
\usepackage{bm}

\begin{document}

\def\Rg{$R_\gamma(z_\gamma>0.4)=0.070\pm0.002\pm0.019\pm0.011$} 


\message{BULLETIZE/SECTIONIZE ALL BACKGROUNDS}


\preprint{CLNS 09/2053}       
\preprint{CLEO 09-06}         

\title{\boldmath Inclusive Radiative $\psi$(2S) Decays}

\author{J.~Libby}
\author{L.~Martin}
\author{A.~Powell}
\author{C.~Thomas}
\author{G.~Wilkinson}
\affiliation{University of Oxford, Oxford OX1 3RH, UK}
\author{H.~Mendez}
\affiliation{University of Puerto Rico, Mayaguez, Puerto Rico 00681}
\author{J.~Y.~Ge}
\author{D.~H.~Miller}
\author{I.~P.~J.~Shipsey}
\author{B.~Xin}
\affiliation{Purdue University, West Lafayette, Indiana 47907, USA}
\author{G.~S.~Adams}
\author{D.~Hu}
\author{B.~Moziak}
\author{J.~Napolitano}
\affiliation{Rensselaer Polytechnic Institute, Troy, New York 12180, USA}
\author{K.~M.~Ecklund}
\affiliation{Rice University, Houston, Texas 77005, USA}
\author{Q.~He}
\author{J.~Insler}
\author{H.~Muramatsu}
\author{C.~S.~Park}
\author{E.~H.~Thorndike}
\author{F.~Yang}
\affiliation{University of Rochester, Rochester, New York 14627, USA}
\author{M.~Artuso}
\author{S.~Blusk}
\author{S.~Khalil}
\author{R.~Mountain}
\author{K.~Randrianarivony}
\author{T.~Skwarnicki}
\author{S.~Stone}
\author{J.~C.~Wang}
\author{L.~M.~Zhang}
\affiliation{Syracuse University, Syracuse, New York 13244, USA}
\author{G.~Bonvicini}
\author{D.~Cinabro}
\author{A.~Lincoln}
\author{M.~J.~Smith}
\author{P.~Zhou}
\author{J.~Zhu}
\author{}
\affiliation{Wayne State University, Detroit, Michigan 48202, USA}
\author{P.~Naik}
\author{J.~Rademacker}
\affiliation{University of Bristol, Bristol BS8 1TL, UK}
\author{D.~M.~Asner}
\author{K.~W.~Edwards}
\author{J.~Reed}
\author{A.~N.~Robichaud}
\author{G.~Tatishvili}
\author{E.~J.~White}
\affiliation{Carleton University, Ottawa, Ontario, Canada K1S 5B6}
\author{R.~A.~Briere}
\author{H.~Vogel}
\affiliation{Carnegie Mellon University, Pittsburgh, Pennsylvania 15213, USA}
\author{P.~U.~E.~Onyisi}
\author{J.~L.~Rosner}
\affiliation{University of Chicago, Chicago, Illinois 60637, USA}
\author{J.~P.~Alexander}
\author{D.~G.~Cassel}
\author{R.~Ehrlich}
\author{L.~Fields}
\author{R.~S.~Galik}
\author{L.~Gibbons}
\author{R.~Gray}
\author{S.~W.~Gray}
\author{D.~L.~Hartill}
\author{B.~K.~Heltsley}
\author{D.~Hertz}
\author{J.~M.~Hunt}
\author{J.~Kandaswamy}
\author{D.~L.~Kreinick}
\author{V.~E.~Kuznetsov}
\author{J.~Ledoux}
\author{H.~Mahlke-Kr\"uger}
\author{J.~R.~Patterson}
\author{D.~Peterson}
\author{D.~Riley}
\author{A.~Ryd}
\author{A.~J.~Sadoff}
\author{X.~Shi}
\author{S.~Stroiney}
\author{W.~M.~Sun}
\author{T.~Wilksen}
\affiliation{Cornell University, Ithaca, New York 14853, USA}
\author{J.~Yelton}
\affiliation{University of Florida, Gainesville, Florida 32611, USA}
\author{P.~Rubin}
\affiliation{George Mason University, Fairfax, Virginia 22030, USA}
\author{N.~Lowrey}
\author{S.~Mehrabyan}
\author{M.~Selen}
\author{J.~Wiss}
\affiliation{University of Illinois, Urbana-Champaign, Illinois 61801, USA}
\author{M.~Kornicer}
\author{R.~E.~Mitchell}
\author{M.~R.~Shepherd}
\author{C.~M.~Tarbert}
\affiliation{Indiana University, Bloomington, Indiana 47405, USA }
\author{D.~Besson}
\affiliation{University of Kansas, Lawrence, Kansas 66045, USA}
\author{T.~K.~Pedlar}
\author{J.~Xavier}
\affiliation{Luther College, Decorah, Iowa 52101, USA}
\author{D.~Cronin-Hennessy}
\author{K.~Y.~Gao}
\author{J.~Hietala}
\author{T.~Klein}
\author{R.~Poling}
\author{P.~Zweber}
\affiliation{University of Minnesota, Minneapolis, Minnesota 55455, USA}
\author{S.~Dobbs}
\author{Z.~Metreveli}
\author{K.~K.~Seth}
\author{B.~J.~Y.~Tan}
\author{A.~Tomaradze}
\affiliation{Northwestern University, Evanston, Illinois 60208, USA}
\collaboration{CLEO Collaboration}
\noaffiliation

\date{\today}

\begin{abstract}
Using $e^+e^-$ collision 
data taken with the CLEO-c detector at the Cornell Electron Storage
Ring, we have investigated the direct photon spectrum in the decay
$\psi({\rm 2S})\to \gamma gg$.
We determine 
the ratio of the inclusive direct photon
decay rate to that of the dominant 
three-gluon decay rate $\psi({\rm 2S})\to ggg$ 
($R_\gamma\equiv \Gamma(\gamma gg)/\Gamma(ggg)$) to
be \Rg, with $z_\gamma$ defined
as the scaled photon energy relative to the
beam energy.
The errors shown are
statistical, systematic, and that due to the uncertainty in
the input branching fractions used to extract the ratio, respectively. This 
$R_\gamma$ value is approximately 2/3 of the comparable value
for the $J/\psi$, both in the region
$z_\gamma>$0.4 and also when extrapolated to the full range of $0<z_\gamma<1$.
\end{abstract}
\pacs{13.20.Gd,13.20.-v,13.40.Hq}
\maketitle

\section{Introduction}
Theoretical approaches to heavy quarkonia radiative decays
were developed shortly after the discovery of charmonium~\cite{r:BCDH78}.
Predictions for the direct photon energy spectrum of
quarkonia were originally
based on 
decays of orthopositronium 
into three photons, leading to the expectation that
the direct photon momentum spectrum should rise linearly with 
$z_\gamma~(\equiv E_\gamma/E_{\rm beam})$ to the kinematic
limit ($z_\gamma\to 1$).
Phase space considerations lead to a slight enhancement 
exactly at the kinematic limit~\cite{r:lowest-qcd,r:Brod-Lep-Mack}. 
Garcia and Soto (GS)~\cite{r:GarciaSoto07} have
performed the most recent calculation of the 
expected direct photon spectrum in charmonium
decays, using an approach similar to that
applied by the same authors for the case of 
$\Upsilon$(1S)$\to \gamma gg$~\cite{GSU1S}.
They model the
endpoint region by combining Non-Relativistic QCD
(NRQCD) with Soft Collinear-Effective Theory (SCET),
which facilitates calculation of the spectrum of the collinear gluons
resulting as $z_\gamma\to 1$. At lower momenta (defined
as $z_\gamma\lesssim$0.4), the
so-called `fragmentation' photon component, due to photon
radiation off final-state quarks, must be
directly calculated. In general (unfortunately), the 
intermediate $z_\gamma$ regime
over which calculations are considered reliable for the $\psi$(2S) 
overlaps poorly with the
higher $z_\gamma$ kinematic regime to which we have the best experimental 
sensitivity.

Since the quantum numbers of the gluon system allow both
color-octet and color-singlet contributions, these must
both be explicitly calculated and summed. 
Of particular relevance to this analysis is the possibility that,
in the context of the GS calculation, the
contribution due to the
octet matrix element can result in a suppression of
the ratio of direct-photon to three-gluon widths (``$R_\gamma$'') for
the $\psi$(2S) compared to the $J/\psi$.
Since the NRQCD matrix element 
appears in the
denominator of $R_\gamma$ and not in the 
numerator (up to order $v^2$ in quark momentum),
a larger value of this matrix element leads to a reduced value of $R_\gamma$.

For direct radiative decays, given a particular photon energy,
theory prescribes
the angular distribution $dN/d\cos\theta_\gamma$ 
for the decay of a vector into three massless vectors
(with $\theta_\gamma$ defined as the polar angle relative to the $e^+$ beam axis).
K\"oller and Walsh considered the angular
spectrum in detail~\cite{r:kol-walsh}, 
demonstrating that, if the
parent is polarized along the beam axis, as the
momentum of the
most energetic primary (photon or gluon)
in $c{\overline c}\to\gamma gg$ or $c{\overline c}\to ggg$ 
approaches the beam energy, the event axis 
increasingly tends to
align with the beam axis. They parametrized the
angular distribution as
$dN/d\cos\theta_\gamma\propto 1+\alpha(z_\gamma)\cos^2\theta_\gamma$;
as $z_\gamma\to$1, $\alpha\to$1.
We note that, according to the
K\"oller-Walsh prescription, the value of
$\alpha$ for intermediate $z_\gamma$ values, where most of the
events occur, is relatively small ($\approx$0.2). Only for $z_\gamma\ge0.9$ is the
forward peaking of the photon angular distribution noticeable.

Although inclusive radiative decays of 
radially excited bottomonia have been 
experimentally investigated~\cite{r:shawn-CLEOIII},
there have been no corresponding measurements for the case of charmonium.
That one previous measurement of the direct photon spectrum in
excited bottomonia ($\Upsilon$(2S) and $\Upsilon$(3S)) found
good agreement in both shape and normalization
with the comparable spectrum
measured for the $\Upsilon$(1S) ground state.

\section{\label{sect:criteria}Data Sets and Event Selection}
The CLEO-c detector is essentially identical to the previous
CLEO~III detector, with the exception of a modified
innermost charged particle tracking system.
Elements of the detector, as well as performance characteristics,
are described in detail elsewhere  \cite{r:CLEO-II,r:CLEOIIIa,r:CLEOIIIb}. 
Over the kinematic regime of
interest to this analysis, the
electromagnetic shower energy resolution is approximately 2\%.
The tracking system, 
RICH particle identification system, 
and electromagnetic
calorimeter are all contained within a 1 Tesla superconducting
solenoid.

The primary data used in this analysis consist of 27.4M 
events collected on the $\psi$(2S) resonance at a
center-of-mass energy $E_{cm}$=3686 MeV, over three distinct running
periods.
To reduce uncertainties from
event-finding efficiencies, we require a minimum of one
charged track found in the event. 
Except for this
charged multiplicity requirement, 
all event selection criteria, photon definitions,
and the generation of background estimators are identical to those 
detailed in our previous analyses of $\Upsilon$~\cite{r:shawn-CLEOIII} and $J/\psi$~\cite{r:Zpsi1S}
data. We therefore present only an abbreviated discussion below.

Events containing two tracks loosely
consistent with leptons and identified topologically
as (radiative) Bhabha or
(radiative) muon pair events are 
explicitly rejected from consideration as candidate
radiative decays, as are events consistent with ``1 vs.\ 3''
$\tau$-pair production.
We additionally reject events 
which have a single, well-identified electron or muon charged track.
Photons are defined as showers detected in the 
barrel ($|\cos\theta_\gamma|\le$0.8) electromagnetic
calorimeter with energy deposition characteristics
consistent with those expected for true photons,
and which are well-isolated from both charged tracks as well
as other showers. 

\section{Analysis}
To obtain $R_{\gamma}$, we must determine separately 
the number of direct photon
events and the number of three gluon events. 
Since approximately 4/5 of all $\psi$(2S) decays are either hadronic or radiative transitions to lower-mass charmonium states, the relative signal to background (primarily due to $\pi^0\to\gamma\gamma$ and $\eta\to\gamma\gamma$) is therefore 
expected to be 
considerably reduced compared to the corresponding $J/\psi$ analysis. Moreover, there is a large direct photon background from the $J/\psi$ itself, which is determined from data and directly subtracted. 
Note that, for the $\psi$(2S), 
there is also a high-energy enhancement
due to radiative continuum processes, such as $e^+e^-\to\gamma\rho$,
which were not present in the 
tagged $J/\psi$ analysis. Continuum contributions
were circumvented in the previous 
$J/\psi\to\gamma gg$ analysis~\cite{r:Zpsi1S} 
using the `tagged' process
$\psi$(2S)$\to\pi\pi J/\psi$; $J/\psi\to\gamma$X.
This continuum contribution
(including initial state radiation [``CO ISR'']) to the observed inclusive 
photon energy spectrum at $E_{cm}$=3686 MeV
is directly determined using data taken at 
center-of-mass energies 15 MeV below the $\psi$(2S) resonance mass.
There is also a correction due to the radiative return to the
$J/\psi$, which subsequently decays into $\gamma gg$, and is
implicitly included in our continuum subtraction.

\subsection{Background Estimation}
\begin{figure}[htpb]
\centerline{\includegraphics[width=8cm]{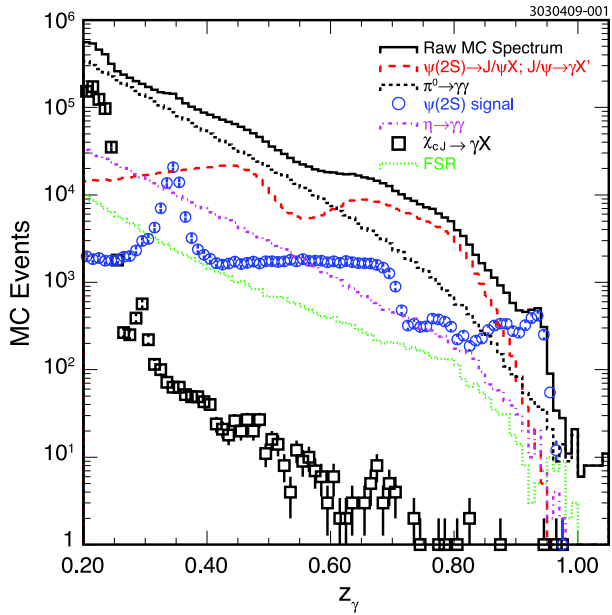}}
\caption{MC simulation of expected backgrounds to direct photon
signal (``FSR'' denotes Final State Radiation).}
\label{fig:psiprNewMCbkgnds}
\end{figure}

Monte Carlo (MC) simulations are used
as a guide to the expected composition of the raw, observed
$E_{cm}$=3686 MeV photon momentum spectrum.
The MC prediction for the
background to the direct photon candidate spectrum, broken
down by parent type and source, is shown in Figure~\ref{fig:psiprNewMCbkgnds}.
The primary background arises from the cascade
process, $\psi$(2S)$\to J/\psi$X; $J/\psi\to\gamma X'$.
This component can be extracted from the data, simply by measuring the inclusive photon spectrum observed
in conjunction with two transition charged pions
having a recoil mass consistent with that of the $J/\psi$.
We scale this observed spectrum up to the total 
number of $\psi$(2S)$\to J/\psi$X events by the 
ratio ${\cal B}(\psi$(2S)$\to J/\psi$X)/(${\cal B}(\psi$(2S)$\to J/\psi\pi^+\pi^-)\epsilon_{\pi^+\pi^-}$), 
with ${\cal B}(\psi$(2S)$\to J/\psi$X)/${\cal B}(\psi$(2S)$\to J/\psi\pi^+\pi^-)=1.784\pm0.003\pm0.02$~\cite{Brianna08}, 
and $\epsilon_{\pi^+\pi^-}$ defined as the reconstruction efficiency for two transition pions, estimated from MC simulations to be 0.63$\pm$0.02. In principle, this cascade photon spectrum depends on the Q-value of the
specific
$\psi$(2S)$\to J/\psi$X transition process; in practice, the smearing of the photon spectrum due to this
variation is negligibly small.

\section{\label{sect:inclusive1s}\boldmath $\pi^0\to\gamma\gamma$ and $\eta\to\gamma\gamma$ Backgrounds}
After subtracting the cascade contribution,
the dominant non-direct photon background arises
primarily from $\pi^0\to\gamma\gamma$ and $\eta\to\gamma\gamma$
decays.
To model the production of $\pi^0$
and $\eta$ daughter photons, two different estimates were
employed in this analysis as well, which we now enumerate. The variation
between the results obtained with these estimators will later
be included in our tabulation of systematic errors.

\subsection{Pseudo-photons}
First, as in our previous $J/\psi$ analysis~\cite{r:Zpsi1S}, we take advantage of the 
expected similar kinematic distributions between charged and
neutral pions, as dictated by isospin symmetry. Although isospin
will break down both when there are decay processes which are not
isospin-symmetric (e.g., $\psi\to\gamma\eta$) or
when the available fragmentation phase space is
comparable to $M_\pi$, in the
intermediate-energy regime (e.g., recoil masses of order 2 GeV), we expect isospin to be reliable. 
For pions produced in strong interactions,
there should therefore
be half as many neutral pions as charged pions. 
We observe agreement
with this expectation to within $\sim$10\% over most of the kinematic regime
relevant for this analysis. 

A ``pseudo-photon'' background spectrum is generated as follows. 
Each charged track passing quality requirements and having
particle identification information consistent with pions is 
decayed isotropically in its rest
frame into two daughter photons, which are then boosted into the lab. 
The photon-finding efficiency from
Monte Carlo simulations is then used to determine the fraction of the
generated pseudo-photons which contribute to the
observed neutral spectrum. Using this procedure,
we not only simulate backgrounds from $\pi^0\rightarrow \gamma \gamma$ decays, 
but also compensate for 
$\eta\rightarrow \gamma \gamma$, by selecting the
appropriate parent mass with a frequency prescribed by MC simulations.

\subsubsection{Cross-check of the pseudo-photon approach}
\begin{figure}[htpb]\centerline{\includegraphics[width=8cm]{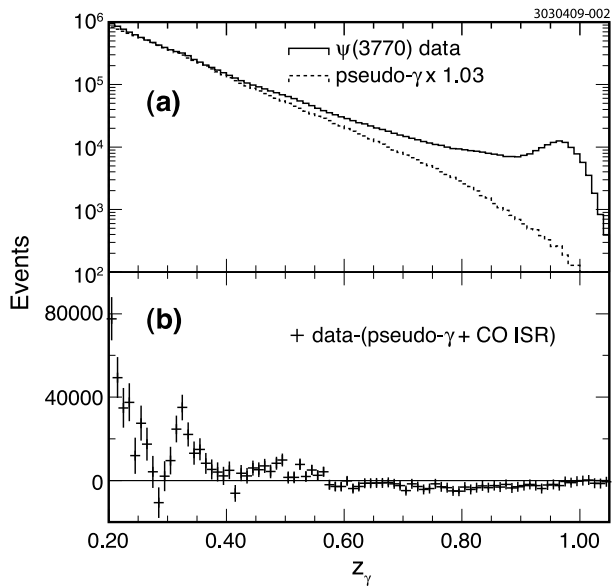}}
\caption{Top (a): $\psi$(3770) data
and pseudo-photon background estimate. Bottom (b): data minus 
both the pseudo-photon and the continuum initial state
radiation (CO ISR) spectra. The 
prominent peak at $z_\gamma\sim0.32$ corresponds to
radiative return to $J/\psi$.}\label{fig:3S2gamX.eps}\end{figure}

In addition to the agreement observed between 
the ``true'' and ``pseudo-photon'' reconstructed
$\pi^0$ yields, as detailed in the $J/\psi$ analysis~\cite{r:Zpsi1S}, our large
$\psi$(3770) sample affords an additional check of
our pseudo-photon approach.
Since the $D{\overline D}$ mode
is kinematically allowed for the $\psi$(3770), the 
direct photon partial width for that resonance should be immeasurably small.
We have therefore compared the inclusive photon spectra from 
the $\psi$(3770) with our estimated pseudo-photon background to determine
how accurately the latter saturates the former. 
Figure~\ref{fig:3S2gamX.eps}(a) shows this direct comparison; Figure~\ref{fig:3S2gamX.eps}(b) shows the residuals
after subtracting both continuum initial
state radiation, and the estimated pseudo-photon background (scaled by 
an empirical factor of 1.03 needed to achieve saturation for $z_\gamma\ge0.4$) 
from the inclusive
$\psi$(3770) spectrum. 
Note the conspicuous presence of the photon line at $z_\gamma\sim$0.32
 in the lower, background-subtracted spectrum, which results
from the radiative return to the
$J/\psi$ resonance.
(Since our continuum data are taken 
at a center-of-mass energy approximately 100 MeV
below the $\psi$(3770) resonance, the radiative 
return peak in the continuum data
will not appear at $z_\gamma\sim$0.32 as it does for the $\psi$(3770) data.)
Aside from that feature, the residuals in the background-subtracted
spectrum are generally featureless. To set the scale, the efficiency-corrected $z_\gamma\ge$0.4 yield after subtraction ($\sim$26K photons) deviates from zero by 1.3\% relative to the initial candidate photon 
count in the same kinematic interval ($\sim$2.1M photons) 
prior to subtraction.
We will later use this 1.03 scale factor as a basis for estimating the
uncertainty in the background normalization for the $\psi$(2S)$\to \gamma gg$ 
analysis.


\subsection{Background estimate from an exponential fit}
As with the $\Upsilon\to \gamma gg$ analysis~\cite{r:shawn-CLEOIII}, we also fit the inclusive photon
spectrum to a smooth exponential curve, in the `control' interval $0.27\le z_\gamma\le0.32$ and extrapolate that
curve into the higher-photon energy region as a background estimator. To the extent that
the control region contains signal, the exponential may result in an over-subtraction
after extrapolation into the high-$z_\gamma$ regime, and therefore, ultimately, an under-estimate of $R_\gamma$.
To assess this bias, we have applied the exponential subtraction technique to Monte Carlo
simulations; we find that the oversubtraction is approximately 9\% for $z_\gamma\ge$0.4. By
comparison, the pseudo-photon subtraction, applied to Monte Carlo simulations, results
in an oversubtraction of less than 5\%. We will later consider these
deviations as implicit in our background systematic uncertainty estimate.

\section{Efficiencies}
\subsection{Photon-finding Efficiency}
In the
barrel region, our photon identification algorithms are approximately 86\% efficient, with
inefficiencies incurred by imposition of a photon isolation requirement and a shower shape
requirement. However, the net photon-finding efficiency is reduced by trigger-
and event-finding requirements, as well as the limited fiducial acceptance of the barrel
calorimeter.

\subsection{Trigger and Event-finding Efficiency}
\begin{figure}[htpb]\centerline{\includegraphics[width=8cm]{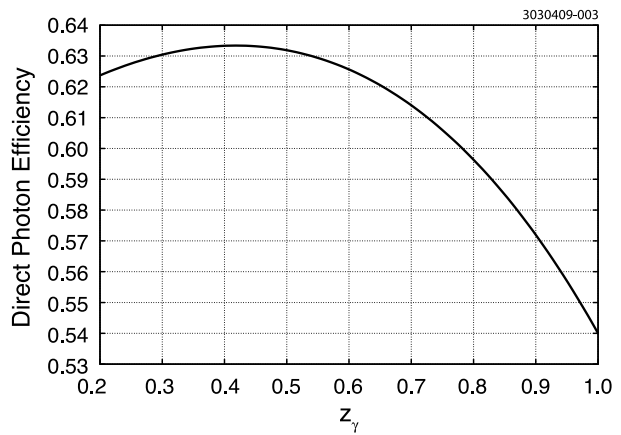}}
\caption{$\psi$(2S)$\to\gamma$X detection efficiency, as a function of scaled photon energy, from MC simulations. Note the suppressed zeroes.}\label{fig:Psi2SEvtCuts.eps}\end{figure}

Based on MC simuations,
the trigger efficiency for events containing a high-energy
photon and at least one charged track is estimated to be 99.3\%.
In general, the pattern of satisfied trigger criteria 
for MC simulations 
shows adequate agreement with data.
The largest
event-selection inefficiencies are incurred by the requirements that
the event
pass the lepton veto, and that the event contain at least one 
well-measured charged track.
Figure~\ref{fig:Psi2SEvtCuts.eps} presents our ``smoothed'' net direct photon
reconstruction efficiency, as a function of $z_\gamma$.
 Loss of efficiency at low $z_\gamma$ is attributed to higher multiplicity events having higher likelihood of one track being mistakenly labeled a lepton. Loss of efficiency at high $z_\gamma$ is attributed to 
reduced charged multiplicity recoiling against high-energy photons 
(and therefore greater likelihood of failing the minimum charged
multiplicity requirement), as well as increased forward-peaking of photons towards the beampipe.

\section{Fits and Signal Extraction}
\begin{figure}[htpb]
\centerline{\includegraphics[width=8cm]{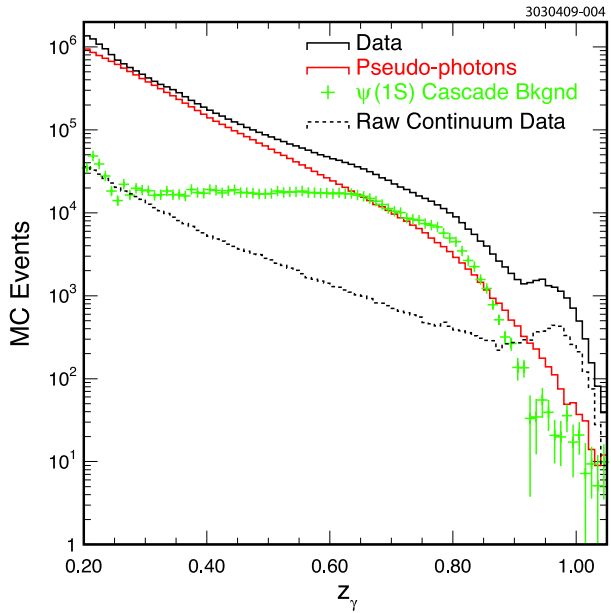}}
\caption{Raw photon momentum spectrum, $\psi$(2S) data, and background estimated using pseudo-photon subtraction scheme.}
\label{fig:psipr-dkpigg-alloverlay}
\end{figure}

After imposing our event selection and photon selection criteria, we determine
the yield in the direct photon energy spectrum. Given the limited statistics, we do not
perform a two-dimensional analysis in both 
$z_\gamma$ and $\cos\theta_\gamma$, as was possible for the $J/\psi$ analysis.
Figure~\ref{fig:psipr-dkpigg-alloverlay} 
shows the raw data, and the components used to extract
the signal, for the pseudo-photon subtraction scheme. 

\begin{figure}[htpb]\centerline{\includegraphics[width=8cm]{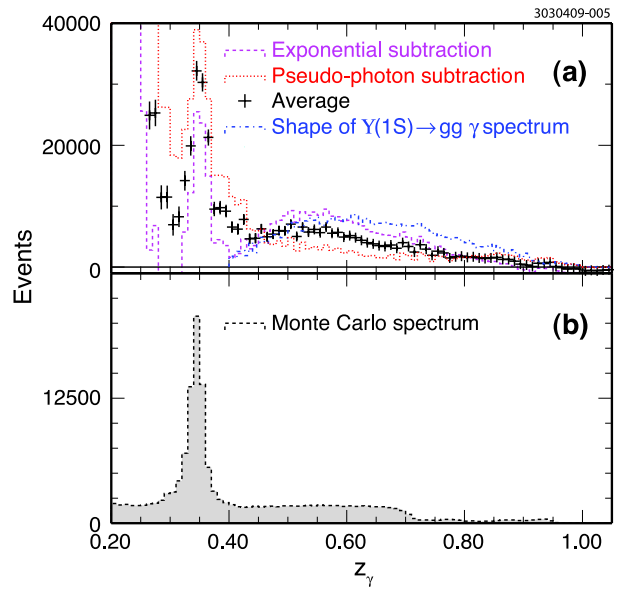}}
\caption{(a) $\psi$(2S)$\to\gamma$X signal spectra, resulting after applying different subtraction schemes. (b) Monte Carlo simulated $\psi$(2S)$\to\gamma$X signal spectrum.}\label{fig:plot-all.eps}\end{figure}

Figure~\ref{fig:plot-all.eps} overlays the signal resulting from the
two different subtraction schemes.
Our final numerical results are based on this Figure.
The total yield is obtained by integrating the point-to-point yields, and adding the individual point-to-point errors in quadrature to obtain the quoted statistical error.
Also shown in Figure~\ref{fig:plot-all.eps}(b) is the MC signal
spectrum, which (given the absence of any real physics in the simulation) 
would not necessarily be expected to show consistency with data. The MC
spectrum does, nevertheless, illustrate the relative scale of our inclusive
signal yield relative to the prominent $\psi$(2S)$\to\gamma\eta_c$ peak at
$z_\gamma\sim$0.34.

\subsection{Comment on Exclusive Modes}
Excluding electromagnetic transitions to other charmonium states (e.g., $\psi$(2S)$\to\gamma\eta_c$), 
two-body radiative exclusive modes should, in principle, be included in our total $\gamma gg$ rate. Although individual
photon lines cannot be resolved in our analysis,
we note that this analysis yields a branching fraction estimate for
$\psi$(2S)$\to\gamma$X ($M_X\le$1.5 GeV) consistent with the tabulated
Particle Data Group (PDG)~\cite{PDG} world average.

\subsection{Extraction of $R_\gamma$}
\subsubsection{Direct Ratio Determination}
To determine $R_\gamma$, we first directly take the ratio:
$(N_{\gamma gg}^{obs}/\epsilon_{\gamma gg})/N_{ggg}$.
The number of $ggg$ events is obtained knowing the
effective $\psi$(2S) cross-section, as tabulated elsewhere~\cite{Brianna08} and subtracting off the
branching fractions to $\chi_{c,J}$, $J/\psi$, $\eta_c$, 
dileptons~\cite{Brianna08}, $q{\overline q}$, and also the
sought-after direct-photon branching fraction (requiring an iteration).
We estimate $2.9\times 10^6$ three-gluon events in our sample.

\subsubsection{\boldmath Normalization to $\psi$(2S)$\to\gamma\eta_c$}
Prominent in both the Monte Carlo simulations and data
points in Figure~\ref{fig:plot-all.eps} is the photon line due to
$\psi$(2S)$\to\gamma\eta_c$.
We can also obtain an estimate of $R_\gamma$ by normalizing the observed photon
yield for $z_\gamma\ge0.4$ to the photon yield observed
corresponding to the decay $\psi$(2S)$\to\gamma\eta_c$, peaking in the interval
$0.32\le z_\gamma\le0.38$. This photon yield is obtained by fitting the
putative $\psi$(2S)$\to\gamma\eta_c$ line to a 
double Gaussian
photon signal (based on the signal shape observed in \cite{Ryan08})
and a smooth lower-order Chebyshev polynomial background.
In this case, the absolute-photon finding largely
cancels; we must correct, however, by the different angular distributions
for the processes $\psi$(2S)$\to\gamma\eta_c$ and 
$\psi$(2S)$\to \gamma gg$, since the latter is 
less forward-peaked than for the
$\eta_c$ transition. Defining $\epsilon_{\cos\theta}$ as the
angular acceptance for $\psi$(2S)$\to\gamma\eta_c$ relative to 
$\psi$(2S)$\to \gamma gg$ ($\approx$0.88),
the value $R_\gamma$ obtained by normalizing to the observed
number $N_{\gamma\eta_c}^{obs}$ of $\psi$(2S)$\to\gamma\eta_c$ events 
can therefore be written as ${\cal B}(\psi$(2S)$\to\gamma\eta_c)\times(N_{\gamma gg}^{obs}/(N_{\gamma\eta_c}^{obs}/\epsilon_{\cos\theta}))/{\cal B}(\psi$(2S)$\to ggg$), with
${\cal B}(\psi$(2S)$\to ggg$) the fraction of $\psi$(2S) events decaying via the three-gluon mode, and
${\cal B}(\psi$(2S)$\to\gamma\eta_c)=(4.3\pm0.6)\times 10^{-3}$~\cite{Ryan08}. We note that
there is an inherent uncertainty of 
$\approx$14\% in this value, resulting from the 
limited statistical precision of the prior CLEO
${\cal B}(\psi$(2S)$\to\gamma\eta_c)$ measurement. 

Numerical inputs used to derive the desired ratio $R_\gamma$ are presented in
Table \ref{tab:calculations}. For internal consistency, we use the most recent CLEO values for $\psi$(2S)$\to J/\psi$X transition branching fractions.
In the last column, we list the explicit source of the
ingredient branching fractions. 
The CLEO-2008 results are from Ref.~\cite{Brianna08}, the PDG
averages are from Ref.~\cite{PDG}, and the CLEO result for
$\psi$(2S)$\to\gamma\eta_c$ is from Ref.~\cite{Ryan08}.
\begin{table*}[htpb] 
\caption{Inputs to calculations and
extracted signal results.
Note that the branching fractions to the $\chi_{cJ}$ states have had
cascades to the $J/\psi$ ($\chi_{cJ}\to J/\psi\gamma$)
removed to avoid double-counting, as indicated.}
\label{tab:calculations}
\begin{tabular}{ccc}  \hline \hline
Quantity & Value & Comment \\ \hline
Number $\psi$(2S) decays produced & 27.4 M & CLEO-2008 \\
${\cal B}(\psi$(2S)$\to\pi^+\pi^-J/\psi$) & ($35.4\pm0.5$)\% & CLEO-2008 \\
${{\cal B}(\psi({\rm 2S})\to J/\psi{\rm X})}/{{\cal B}(\psi({\rm 2S})\to\pi^+\pi^-J/\psi)}$ & $1.784\pm0.003\pm0.02$ & CLEO-2008 \\
${\cal B}(\psi$(2S)$\to e^+e^-)$ & ($0.765\pm0.017)$\% & PDG08 \\
${\cal B}(\psi$(2S)$\to \mu^+\mu^-)$ & ($0.76\pm0.08$)\% & PDG08 \\
${\cal B}(\psi$(2S)$\to \tau^+\tau^-)$  & ($0.3\pm0.04$)\% & PDG08 \\
${\cal B}(\psi$(2S)$\to\gamma^*\to~hadrons)$ & ($1.75\pm0.14$)\% & PDG08 \\
${\cal B}(\psi$(2S)$\to\gamma\chi_{c0})\times(1-{\cal B}(\chi_{c0}\to\gamma J/\psi))$ & $[(9.42\pm0.31)\times$0.99]\% & PDG08 \\
${\cal B}(\psi$(2S)$\to\gamma\chi_{c1})\times(1-{\cal B}(\chi_{c1}\to\gamma J/\psi))$ & [$(9.2\pm0.4)\times$0.66]\% & PDG08 \\
${\cal B}(\psi$(2S)$\to\gamma\chi_{c2})\times(1-{\cal B}(\chi_{c2}\to\gamma J/\psi))$ & [$(8.69\pm0.35)\times$0.8]\% & PDG08 \\ 
${\cal B}(\psi$(2S)$\to\gamma\eta_c)$ & ($0.43\pm0.06$)\% & CLEO \\
$f_z$ & 0.725 & $z_\gamma\ge$0.4 fraction of entire $\gamma gg$ spectrum \\ 
$N_{ggg}$ & 2.9 M & \\ \hline
$N_\gamma(z_\gamma\ge0.4)$ & 162 K & pseudo-photon subtraction \\
$N_\gamma(z_\gamma\ge0.4)$ & 232 K & exponential subtraction \\
$R_\gamma(z_\gamma\ge0.4)$ & ($5.65\pm0.03$)\% & pseudo-photon subtraction \\
$R_\gamma(z_\gamma\ge0.4)$ & ($8.34\pm0.04$)\% & exponential subtraction \\ \hline
$N_{\gamma\eta_c}$ & 130 K & from fit to photon line (averaged plot) \\ 
$R_\gamma(z_\gamma\ge0.4)$ & ($7.07\pm0.8$)\% & normalized to ${\cal B}(\psi$(2S)$\to\gamma\eta_c)$  \\ \hline \hline
\end{tabular}
\end{table*}
Averaging the last three $R_\gamma$ lines in Table \ref{tab:calculations}, we obtain
\Rg, with statistical, systematic,
and branching-ratio uncertainties shown, limited to $z_\gamma\ge$0.4.

\section{Systematic errors}
We identify and estimate systematic errors in our $R_\gamma$
determination as follows:
\begin{enumerate}
\item Given the fact that the signal-to-noise
ratio is so small, it is important to 
evaluate the sensitivity to the background scale.
We have correspondingly evaluated the change in $R_\gamma$ when the
background estimators are individually
toggled in normalization
by 3\% (typical of the normalization uncertainty in previous $\Upsilon$ and
$\psi$ analyses and consistent with Fig.~\ref{fig:3S2gamX.eps}). For $\pm$3\% variation of the pseudo-photon and exponential
normalizations, we find fractional $R_\gamma$ variations of order 20\%. We
can independently assess the uncertainty in the background normalization by
taking the difference between the exponential and
pseudo-photon extracted $R_\gamma$ values (19\%). Conservatively
taken together, we
quote a total background normalization uncertainty of 27\%. This
is by far our largest single systematic error.
\item Extrapolation of direct photon spectrum to $z_\gamma\le$0.4: Although this
obviously does not enter directly into the calculated yield for $z_\gamma\ge$0.4,
this quantity does enter in the calculation of $R_\gamma$. 
Since $R_\gamma$ is defined relative to the three-gluon yield, we must subtract the entire
estimated $\gamma gg$ rate, including the unseen portion below $z_\gamma=0.4$,
to estimate the fraction of direct $\psi$(2S) decays
proceeding through $ggg$-only. We estimate this error as half the
total difference between the fractional yield for
$z_\gamma\le$0.4 expected in an $\Upsilon$(1S)-like direct
photon spectrum (28\%), vs. the extreme case of a linear extrapolation
to $z_\gamma$=0 (40\%), or
6\%.
However, since this enters as second-order
in the value of $R_\gamma$, this contribution to the
total systematic error in
$R_\gamma$ is only $\approx 1$\%.
\item Angular distribution uncertainties: Although we have not performed
a two-dimensional fit, we have compared the obtained values of $R_\gamma$
over high- ($0.4\le|\cos\theta_\gamma|\le0.8$) and low- ($0.4\ge|\cos\theta_\gamma|$) polar angle regimes. We find a difference of 4.4\% 
between these two cases, and
assign a corresponding error of 2.2\%.
\item The contribution to the signal 
due to non-photon showers, based on MC tagging studies,
is estimated to be 4\% (the total of the $K^0_L$ plus
antineutron contribution, as indicated by MC studies)
after applying all cuts. Note
that these contributions are largely subsumed into
our exponential background subtraction. 
\item The trigger efficiency systematic
error in the ratio is $\le$1\%.
\item Photon-finding efficiency uncertainty contribution to the ratio is estimated at 2\%.
\item The uncertainty contribution due to the limited precision of 
${\cal B}(\psi$(2S)$\to J/\psi X)$ is estimated at 2\%.
\item The uncertainty in the three-gluon event-finding efficiency results in
a systematic error of 2\%.
\item In principle, the $J/\psi\to \gamma gg$ cascade photon spectrum, which is calculated
from the observed $\psi$(2S)$\to\psi\pi^+\pi^-$; $J\psi\to\gamma X'$ spectrum, can differ from the cascade spectrum
for different cascade processes, with different typical boosts.
For a mean dipion mass of
400 MeV, the $J/\psi$ recoil momentum is approximately 390 MeV/c; for a cascade transition
through $\chi_{cJ}$ (i.e. $\psi$(2S)$\to\gamma\chi_{cJ}$; $\chi_{cJ}\to\gamma J/\psi$),
assuming that the two transition photons are uncorrelated in angle, the average daughter 
$J/\psi$ momentum is about 10 MeV/c higher. For our purposes, we consider this effect
negligible. 
\item We include (separately) an error of 15\% based on the difference in results obtained using CLEO-only values vs. the most recently compiled values from the Particle Data Group as branching fraction inputs.
\end{enumerate}

Systematic errors are summarized in Table \ref{tab:sys}.
\begin{table}[htpb]
\begin{center}
\caption{Systematic Errors. \label{tab:sys}}
\begin{tabular}{cc}\hline \hline
Source                           &$\delta R_\gamma/R_\gamma$ (\%)  \\ \hline 
Background normalization & 27 \\
$z_\gamma\to 0$ extrapolation & 1 \\
Angular distribution uncertainties & 2.2 \\
$K^0_L$ + anti-neutron contamination & 4 \\ 
Trigger efficiency & 1 \\
Photon-finding & 2 \\
$\psi$(2S)$\to J/\psi$X normalization & 2 \\
Three-gluon event efficiency & 2 \\ \hline
Input branching fractions & 15 \\ \hline \hline
{\bf Total Systematic Error} & $\pm27\pm15$ \\ \hline \hline
\end{tabular}
\end{center}
\end{table}

\section{\boldmath Comparison to $J/\psi$}
Our expectation is
that the ratio of partial widths, $\Gamma(\gamma gg)/\Gamma(ggg)$ should be the
same for the $\psi$(2S) as for the $J/\psi$. This expectation is satisfied for
the case of the $\Upsilon$ system.
We have calculated $R_\gamma(\psi(2S))/R_\gamma(J/\psi)$, although
we note that since
the two signal extraction techniques are very different, the dominant 
systematic errors are largely uncorrelated for both numerator and
denominator, so therefore, there is no cancellation of common
systematic errors in this quotient. 
Restricted to the interval $z_\gamma\ge$0.4, the ratio of $\gamma gg$
partial widths for the $\psi$(2S) compared to the $J/\psi$ is 
$0.69\pm0.20$.
Assuming that the direct photon momentum spectral
shape for the $\psi$(2S) is similar to
the $J/\psi$, the fraction of photons in the region $z_\gamma\le0.4$ is
approximately 28\% of the entire spectrum, corresponding to a momentum-integrated
value of $R_\gamma=0.097\pm0.026\pm0.010$, again about 2/3 the
value obtained at the $J/\psi$. (If instead, we extrapolate linearly
from $z_\gamma=0.4$ to $z_\gamma=0.0$, this ratio of widths increases
by approximately 10\%.) 
Although this low value in comparison to the $J/\psi$ may seem
surprising, such apparent anomalies~\cite{GP09} are not uncommon when
ratios of $\psi$(2S)/$J/\psi$ partial widths are taken.
The well-known VP-suppression observed in
$\psi$(2S) exclusive hadronic decays\cite{VP} can be considerably more
dramatic, as is the recently documented dearth of 
radiative $\psi$(2S) decays to $\eta$ and $\eta^\prime$~\cite{GP09}
      relative to those from $J/\psi$.
We note that the ratio of dileptonic to three-gluon widths
($B_{ll}/B_{ggg}$) is also considerably smaller for the
$\psi$(2S) compared to the $J/\psi$. 
\message{Taken together, this 
pattern suggests that the 'problem' may
lie in the 3-gluon denominator common to both ratios.}

\section{Summary}
We have made the first measurement of the direct photon spectrum in
$\psi$(2S) decays, obtaining an inclusive rate (relative to 
three-gluon decays) approximately 2/3 that obtained at the
$J/\psi$, and suggesting larger $\psi$(2S) corrections, possibly due to 
closer proximity to $D{\overline D}$ threshold.
Although theoretical predictions for this value are
somewhat scarce, we note that this is now one of 
several apparent anomalies observed in ratios of widths between the 
$\psi$(2S) and the $J/\psi$. 
This study completes the set of $R_\gamma$ measurements
for all bound nS states of heavy quarkonia below open flavor threshold.
It is hoped that
further theoretical study will hopefully elucidate this result.

\section{Acknowledgments}
We gratefully acknowledge the effort of the CESR staff
in providing us with excellent luminosity and running conditions.
Shawn Henderson wrote initial verions of the computer codes used in this
data analysis.
We thank Xavier Garcia i Tormo and Joan Soto for illuminating discussions.
D.~Cronin-Hennessy and A.~Ryd thank the A.P.~Sloan Foundation.
This work was supported by the National Science Foundation,
the U.S. Department of Energy,
the Natural Sciences and Engineering Research Council of Canada, and
the U.K. Science and Technology Facilities Council.


\end{document}